\documentclass[twocolumn,11pt]{article}
\usepackage{amssymb,amsmath,latexsym}

\usepackage[square,numbers]{natbib}
\usepackage{chemarr}
\usepackage{enumitem}
\usepackage{graphicx}
\usepackage{amssymb}
\usepackage{lineno}
\usepackage{amsmath}
\usepackage{multirow}
\usepackage{float}
\begin{document}
\date{\today}
\newcommand{\MSun}{{M_\odot}}
\newcommand{\LSun}{{L_\odot}}
\newcommand{\Rstar}{{R_\star}}
\newcommand{\calE}{{\cal{E}}}
\newcommand{\calM}{{\cal{M}}}
\newcommand{\calV}{{\cal{V}}}
\newcommand{\calO}{{\cal{O}}}
\newcommand{\calH}{{\cal{H}}}
\newcommand{\calD}{{\cal{D}}}
\newcommand{\calB}{{\cal{B}}}
\newcommand{\calK}{{\cal{K}}}
\newcommand{\labeln}[1]{\label{#1}}
\newcommand{\Lsolar}{L$_{\odot}$}
\newcommand{\farcmin}{\hbox{$.\mkern-4mu^\prime$}}
\newcommand{\farcsec}{\hbox{$.\!\!^{\prime\prime}$}}
\newcommand{\kms}{\rm km\,s^{-1}}
\newcommand{\cc}{\rm cm^{-3}}
\newcommand{\Alfven}{$\rm Alfv\acute{e}n$}
\newcommand{\Vap}{V^\mathrm{P}_\mathrm{A}}
\newcommand{\Vat}{V^\mathrm{T}_\mathrm{A}}
\newcommand{\D}{\partial}
\newcommand{\DD}{\frac}
\newcommand{\TAW}{\tiny{\rm TAW}}
\newcommand{\mm }{\mathrm}
\newcommand{\Bp }{B_\mathrm{p}}
\newcommand{\Bpr }{B_\mathrm{r}}
\newcommand{\Bpz }{B_\mathrm{\theta}}
\newcommand{\Bt }{B_\mathrm{T}}
\newcommand{\Vp }{V_\mathrm{p}}
\newcommand{\Vpr }{V_\mathrm{r}}
\newcommand{\Vpz }{V_\mathrm{\theta}}
\newcommand{\Vt }{V_\mathrm{\varphi}}
\newcommand{\Ti }{T_\mathrm{i}}
\newcommand{\Te }{T_\mathrm{e}}
\newcommand{\rtr }{r_\mathrm{tr}}
\newcommand{\rbl }{r_\mathrm{BL}}
\newcommand{\rtrun }{r_\mathrm{trun}}
\newcommand{\thet }{\theta}
\newcommand{\thetd }{\theta_\mathrm{d}}
\newcommand{\thd }{\theta_d}
\newcommand{\thw }{\theta_W}
\newcommand{\beq}{\begin{equation}}
\newcommand{\eeq}{\end{equation}}
\newcommand{\ben}{\begin{enumerate}}
\newcommand{\een}{\end{enumerate}}
\newcommand{\bit}{\begin{itemize}}
\newcommand{\eit}{\end{itemize}}
\newcommand{\barr}{\begin{array}}
\newcommand{\earr}{\end{array}}
\newcommand{\bc}{\begin{center}}
\newcommand{\ec}{\end{center}}
\newcommand{\DroII}{\overline{\overline{\rm D}}}
\newcommand{\DroI}{{\overline{\rm D}}}
\newcommand{\eps}{\epsilon}
\newcommand{\veps}{\varepsilon}
\newcommand{\vepsdi}{{\cal E}^\mathrm{d}_\mathrm{i}}
\newcommand{\vepsde}{{\cal E}^\mathrm{d}_\mathrm{e}}
\newcommand{\lraS}{\longmapsto}
\newcommand{\lra}{\longrightarrow}
\newcommand{\LRA}{\Longrightarrow}
\newcommand{\Equival}{\Longleftrightarrow}
\newcommand{\DRA}{\Downarrow}
\newcommand{\LLRA}{\Longleftrightarrow}
\newcommand{\diver}{\mbox{\,div}}
\newcommand{\grad}{\mbox{\,grad}}
\newcommand{\cd}{\!\cdot\!}
\newcommand{\Msun}{{\,{\cal M}_{\odot}}}
\newcommand{\Mstar}{{\,{\cal M}_{\star}}}
\newcommand{\Mdot}{{\,\dot{\cal M}}}
\newcommand{\ds}{ds}
\newcommand{\dt}{dt}
\newcommand{\dx}{dx}
\newcommand{\dr}{dr}
\newcommand{\dth}{d\theta}
\newcommand{\dphi}{d\phi}

\newcommand{\pt}{\frac{\partial}{\partial t}}
\newcommand{\pk}{\frac{\partial}{\partial x^k}}
\newcommand{\pj}{\frac{\partial}{\partial x^j}}
\newcommand{\pmu}{\frac{\partial}{\partial x^\mu}}
\newcommand{\pr}{\frac{\partial}{\partial r}}
\newcommand{\pth}{\frac{\partial}{\partial \theta}}
\newcommand{\pR}{\frac{\partial}{\partial R}}
\newcommand{\pZ}{\frac{\partial}{\partial Z}}
\newcommand{\pphi}{\frac{\partial}{\partial \phi}}

\newcommand{\vadve}{v^k-\frac{1}{\alpha}\beta^k}
\newcommand{\vadv}{v_{Adv}^k}
\newcommand{\dv}{\sqrt{-g}}
\newcommand{\fdv}{\frac{1}{\dv}}
\newcommand{\dvr}{{\tilde{\rho}}^2\sin\theta}
\newcommand{\dvt}{{\tilde{\rho}}\sin\theta}
\newcommand{\dvrss}{r^2\sin\theta}
\newcommand{\dvtss}{r\sin\theta}
\newcommand{\dd}{\sqrt{\gamma}}
\newcommand{\ddw}{\tilde{\rho}^2\sin\theta}
\newcommand{\mbh}{M_{BH}}
\newcommand{\dualf}{\!\!\!\!\left.\right.^\ast\!\! F}
\newcommand{\cdt}{\frac{1}{\dv}\pt}
\newcommand{\cdr}{\frac{1}{\dv}\pr}
\newcommand{\cdth}{\frac{1}{\dv}\pth}
\newcommand{\cdk}{\frac{1}{\dv}\pk}
\newcommand{\cdj}{\frac{1}{\dv}\pj}
\newcommand{\rad}{\;r\! a\! d\;}
\newcommand{\half}{\frac{1}{2}}
\newcommand{\ARZL}{\textquotedblleft}
\newcommand{\ARZR}{\textquotedblright}
  \title{ How massive are the superfluid cores in the Crab and Vela pulsars and \\
             why their glitch-events are accompanied with under and overshootings? }

\author{\thanks{E-mail:AHujeirat@uni-hd.de}~Hujeirat  A.A., \thanks{ravi.samtaney@kaust.edu.sa} Samtaney, R. \\
IWR, Universit\"at Heidelberg, 69120 Heidelberg, Germany \\
Applied Mathematics and Computational Science, CEMSE Division, KAUST}
\maketitle
\abstract{

The Crab and Vela are well-studied glitching pulsars and the data obtained sofar should enable us to test the reliability of models
of their internal structures.\\
Very recently it was proposed that glitching pulsars  are embedded in bimetric spacetime:
their incompressible superfluid cores (SuSu-cores) are embedded in  flat spacetime, whereas the ambient compressible and dissipative
media are enclosed in Schwarzschild spacetime. \\
In this letter we apply this model to the Crab and Vela pulsars and show that a newly born pulsar   initially of  $1.25\MSun$  and
an embryonic SuSu-core of $0.029\MSun$ could  evolve   into a \ARZL{Crab-like}\ARZR pulsar after 1000 years and into a
\ARZL{Vela-like}\ARZR pulsar
10000 years later to finally fade away as an invisible dark energy object after roughly 10 Myr.   \\
Based thereon we infer that the Crab and the Vela pulsars should have SuSu-cores of  $0.15\MSun$  and $0.55\MSun,$
respectively.\\
 Futhermore, the under- and overshootings phenomena observed to accompany the glitch events of the Vela pulsar are rather a common phenomenon
 of glitching pulsars that can be well-explained within the framework of bimetric spacetime.\\
    }
    \\

 \section{Observational constraints and methodology}
  The Crab and Vela pulsars are well-known and extensively studied pulsars (see \citep{Roy2012,Yu2013,Yuan2019,Eya2014,Espinoza2011,Fuentes2017,ReviewPulsars16} and the references therein).  In Table (\ref{Table1}) we summarize their basic observational data
    relevant for the present discussion, are summarized. The SuSu-Scenario relies on solving the TOV equation in combination with the equations of torque balance between the incompressible
superfluid core, whose dynamics obey the  Onsager-Feymann equation, and an overlying shell of compressible and dissipative matter
 (see Sec. 2 and Eq. 10 in \cite{HujeiratGlitch18}).
\begin{center}
\begin{table}
\begin{tabular}{ |l|c|c|c| }
\hline
    & Crab & Vela \\
\hline
Mass ($\MSun$)                                & 1.4 & 1.8 \\
Age (kyr)                                            & 1.24 & 11.3 \\
B $(10^{12}\ G)$                               & 4.875 & 4.35 \\
$\Omega\, (s^{-1})$                            & 200 & 70 \\
$\Delta\Omega_g/\Omega$               & $ 4\times 10^{-9}$ & $2.338\times 10^{-6}$ \\
$\Delta t_g (yr)$                                 & 1.6 & 2.5 \\
\hline
\end{tabular}
\caption{\label{Table1} A list of the main observational data of the Crab and Vela pulsars relevant for the
present study  (see \citep{Yu2013,Yuan2019,Eya2014,Espinoza2011,Fuentes2017,ReviewPulsars16} and the references therein).}
\end{table}
\end{center}
\begin{figure}[htb]
\centering {\hspace*{-0.35cm}
\includegraphics*[angle=-0, width=7.0cm]{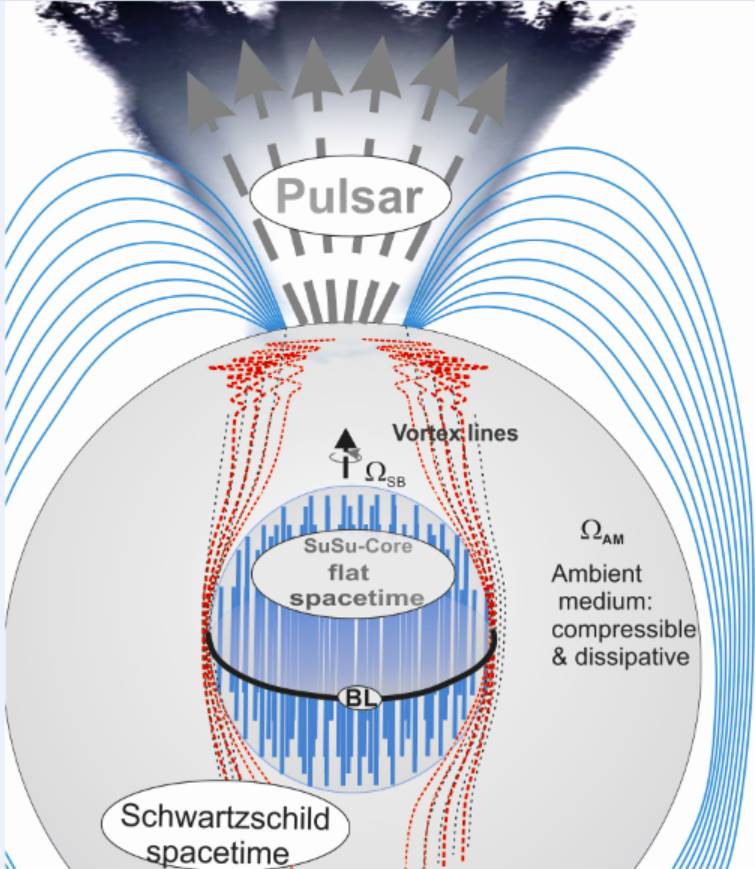}
}
\caption{\small  A schematic description of the bimetric spacetime inside glitching pulsars: The  incompressible superfluid core
is embedded in a Minkowski spacetime whereas the ambient media are enclosed in Schwarzschild spacetime.
}\label{NSInternal}
\end{figure}
In  \cite{HujeiratRavi19} these equations were solved at the background of a bimetric spacetime (see Fig.\ref{NSInternal}). Unlike the original model \cite{HujeiratGlitch18}, in which the spin-down of  the SuSu-core
 is set to follow  an a priori given sequence of values  $\{\Omega_{c}^n\},$  in the present work however, the SuSu-core is set to undergo  an abrupt spin-down, if the difference between its eigen rotation and that of the ambient medium surpasses a critical value $\{\Delta\Omega_{cr}^n\},$ i.e.
  if $ \Delta\Omega^n_{c-am} = \Omega^n_c -  \Omega^n_{am} \geq \Delta\Omega^n_{cr},$  where `n' and `am' refer to  the order of the elements in the relevant sequence and to the ambient medium, respectively. This approach is more consistent than the former, as the elements of  $ \Delta\Omega^n_{cr}$ are determined here
  through the rate of loss of   rotational energy of the entire star and that these should overlap the current values observed in the Crab and Vela pulsars. \\
\begin{figure}[t]
\centering {\hspace*{-0.35cm}
\includegraphics*[angle=-0, width=7.15cm]{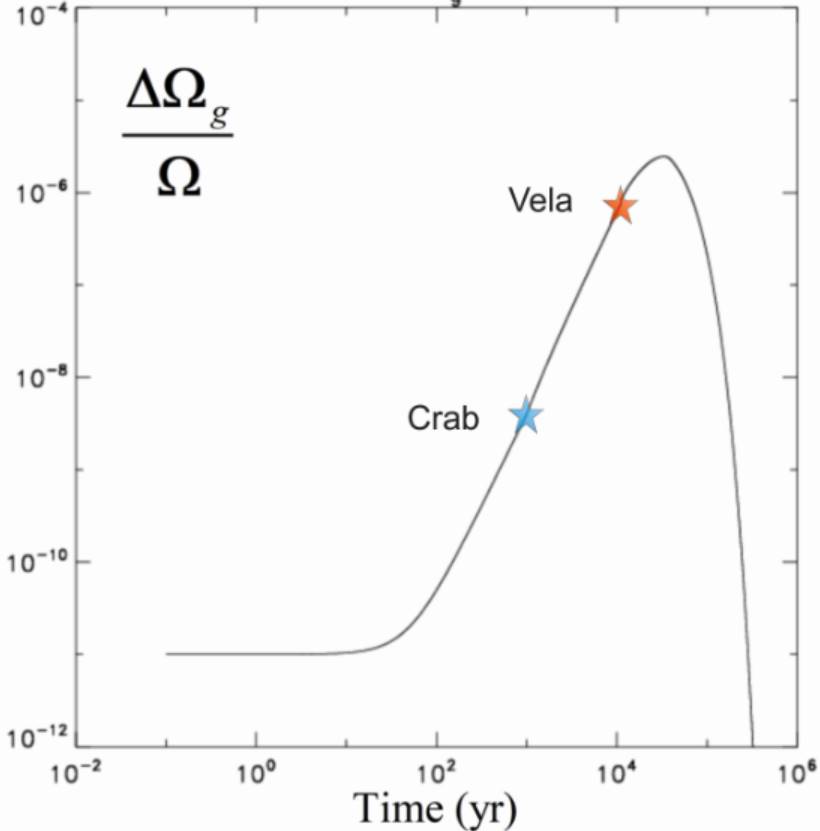}
}
\caption{\small The elements of the quantum sequence  $\Delta \Omega^n_g/\Omega$ are shown as function of time in year units.
Both the core and the ambient medium are set to rotate with the same frequency $1440~s^{-1} initially.$
As the pulsar cools down,  $\Delta \Omega^n_g/\Omega$ starts increasing  to reach  $4\times 10^{-9}$ after approximately 1000 yr,
(which corresponds to the Crab phase/blue-star) and $8.15\times 10^{-7}$ after 11000 yr (which corresponds to Vela phase/red-star).
$\Delta \Omega^n_g/\Omega$ here is measured in units of $\Omega =200$/s.
}
 \label{domg_g}
\end{figure}
 The strategy of obtaining the opitimal values here relies on using a global  iterative solution procedure that takes the
  the following  constraints into account ( see also Table 1):
  \ben
  \item  The elements  of the sequence $ \{ \DD{\Delta\Omega_{g}}{\Omega}\}^n $ must  fulfill the three condions:
  \bit
  \item    $\{\DD{\Delta\Omega_{g}}{\Omega}\}^n   \xrightarrow[n \to 0]{} 0 ,$ which means that the media in both the core and in the surrounding shell
              must have identical rotational frequency initially.
  \item   ${\DD{\Delta\Omega_{g}}{\Omega}}|_{n=N_0} =   {\DD{\Delta\Omega_{g}}{\Omega}}|_{Crab} = 4 \times 10^{-9}$  and \\
  ${\DD{\Delta\Omega_{g}}{\Omega}}|_{n=N_1} =   {\DD{\Delta\Omega_{g}}{\Omega}}|_{Vela} = 2.33 \times 10^{-6},$ i.e.  the elements number $N_0$ and  $N_1( \gg N_0)$ of the sequence $ \{ \DD{\Delta\Omega_{g}}{\Omega}\}^n $ must be identical to the observed values of the Crab and to the
      Vela pulsars, respectively.
   \item   $\{\DD{\Delta\Omega_{g}}{\Omega}\}^n   \xrightarrow[n \to \infty]{} \alpha_\infty < \infty,  $ i.e.
   the sequence  must converges to a finite value. Moreover, our test calculation have shown that
   for $ t\gg t_{Vela},$  $\DD{\D}{\D t} (\{\DD{\Delta\Omega_{g}}{\Omega}\}^n) <0$ as otherwise the magnetic
   field would fail to spin-down the crust and therefore to surpass  $\Delta\Omega_{cr}$ required for triggering a prompt  spin-down of the
   the core  into the next lower energy state.\\
   Indeed, one possible sequence which fulfills the above-mentioned constraints, though it might not be unique, is shown in Fig. (\ref{domg_g}).
  \eit
\begin{figure}[t]
\centering {\hspace*{-0.35cm}
\includegraphics*[angle=-0, width=7.15cm]{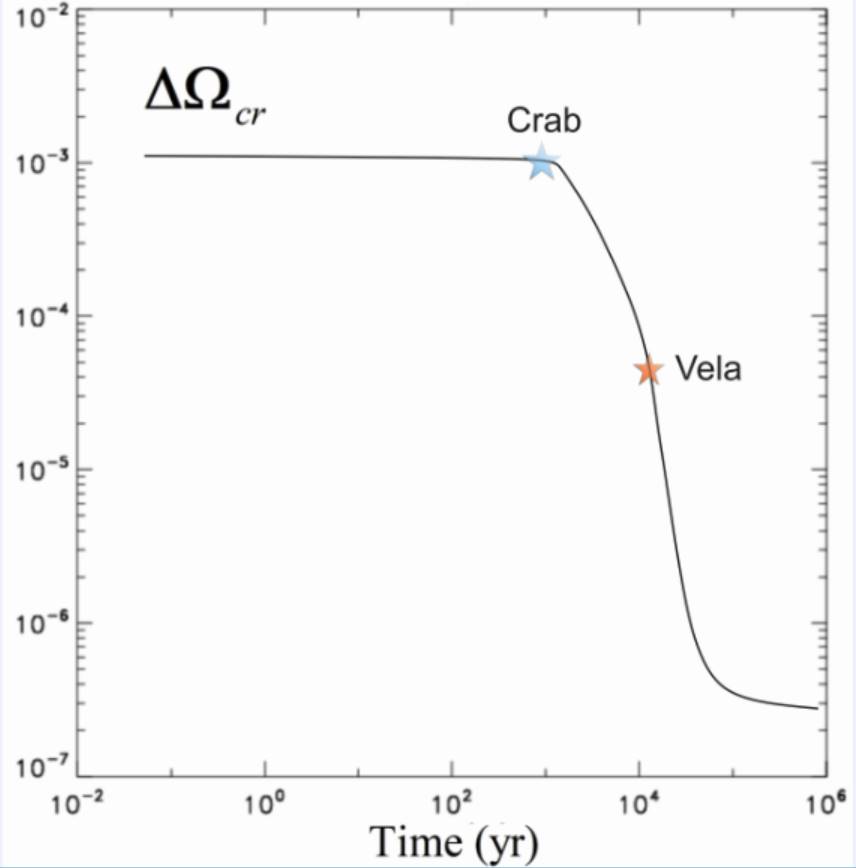}
}
\caption{\small  The elements of the sequence  $\Delta \Omega^n_{cr}$ versus cosmic time. Each element corresponds to the critical
difference between the rotational of frequency of the core and that of the ambient medium, beyond which the core undergoes
 a prompt spin-down to the next lower energy state.
}  \label{domg_cr}
\end{figure}
 \item The initial conditions used here are   $\Omega_0=\Omega(t=0) =   1440$  Hz  (see \cite{Hansel1999} and the references therein).
          Here both the core and the ambiant  medium are set to initially  rotate with the same  frequency, i.e.  $ \Omega^n_c =  \Omega^n_{am}(t=0)=  \Omega_0.$ The initial total mass of the pulsars  and that of the core  are taken to be:
           $M_c= \epsilon_0 \MSun $  and  $M_0 =M(t=0)= \epsilon_1 \MSun, $ respectively,  where $\epsilon_0$ and $ \epsilon_1$ are parameters whose values are determined through a global iteration procedure. The initial magnetic field strength is taken to be  $B(t=0) = 10^{13}$ Gauss.

 \item The elements of the sequence $\{\Omega_{am}\}^n$ are obtained  through the energy balance equation:
        \beq
            \DD{d}{dt} (\DD{1}{\Omega^2_{am}})      = - \alpha_{EM} \DD{B^2}{I_{am}},
        \eeq
        where $I_{am}$ is the inertia of the ambient compressible dissipative medium, which, due to the increase of the SuSu-core, must
          decrease on the cosmic time and $\alpha_{EM} = 4.9\,10^{-4}$ is a non-dimensional constant. \\
          Thus, the time-evolution of the core's rotational frequency proceeds as follows: for a given $\Omega^n_c,$ the ambient medium is set to decrease its frequency continuously  with time through the emission
           magnetic dipole radiation.  This implies that  the difference $\Delta\Omega_{c-am}$ should increase with time until
           $ \Delta \Omega_{c-am} $ has surpassed the critical value $\geq \Delta\Omega_{cr}.$  In this case three events are expected
           to occur promptly:
           \bit
           \item The rigid-body rotating core changes abruptly  its rotational state from
             $E^{n,rot}_c = \DD{1}{2}I^n_c (\Omega^n_c)^2$ into the next the quantum-mechanically permitted lower energy state:
             $E^{n+1,rot}_c = \DD{1}{2}I^{n+1}_c (\Omega^{n+1}_c)^2.$ This process is associated with ejection of a certain number
             of vortices into the boundary layer (BL) between the core and the overlying dissipative medium.
           \item The ejected vortices by the core are then absorbed by the differentially rotating dissipative medium and re-distributed viscously.
                    Hence the medium in the BL would experience the prompt spin-up:
                     $ \Omega^n_{am} \rightarrow \Omega^n_{am} + \Delta\Omega^n_g, $ where $ \Delta\Omega^n_g$ is deduced from the
                     sequence $ \{ \DD{\Delta\Omega_{g}}{\Omega}\}^n .$
           \item  The radius of the core is set to increases as dictated by the Onsager-feynmann equation:
           \beq
           \oint V\cdot d \ell =  \DD{2\pi\hbar }{m} N,
           \eeq
           where $V,~ \ell, ~, \hbar, ~m,~ N, $ denote the velocity vector, the vector of
line-element, the reduced Planck constant, mass of the superfluid particle pair and the number of vortices, respectively (see \cite{HujeiratMassiveNSs18} for further details).
Imposing zero-torque condition on the incompressible SuSu-core, i.e., $\DD{d}{dt}(\Omega I)_c = 0.$ we then  obtain the following recursive relation:
           \beq
            \begin{array}{ll}
             (\Omega S)^{n+1}_c =  (\Omega S)^{n}_c & \\
              \Rightarrow S^{n+1} = (\DD{\Omega^{n}_c}{\Omega^{n+1}_c}) \,S^{n}
           \end{array}
           \eeq
           where  $S^{n}_c\doteq \pi (R^n_c)^2$ and $R^n_c$ correspond to  the the cross-sectional area of the SuSu-core and to the corresponding radius, respectively.  The increase in the dimension of the core implies
            that the matter in the geometrically thin boundary layer between the SuSu-core and the ambient medium should undergo a crossover phase transition into an incompressible superfluid, whose total energy density saturates around the critical value $\rho_{cr} \approx 6 \rho_0$
            (see \cite{HujeiratMetamor18} and the references therin).
            The growth of the core proceeds  on the cosmic time scale and ends when the pulsar has metamorphosed entirely into a maximally compact  invisible dark energy object and therefore becomes observationally indistinguishable from a stellar black hole.
           \eit
 \een

\begin{figure}[h]
\centering {\hspace*{-0.35cm}
\includegraphics*[angle=-0, width=7.15cm]{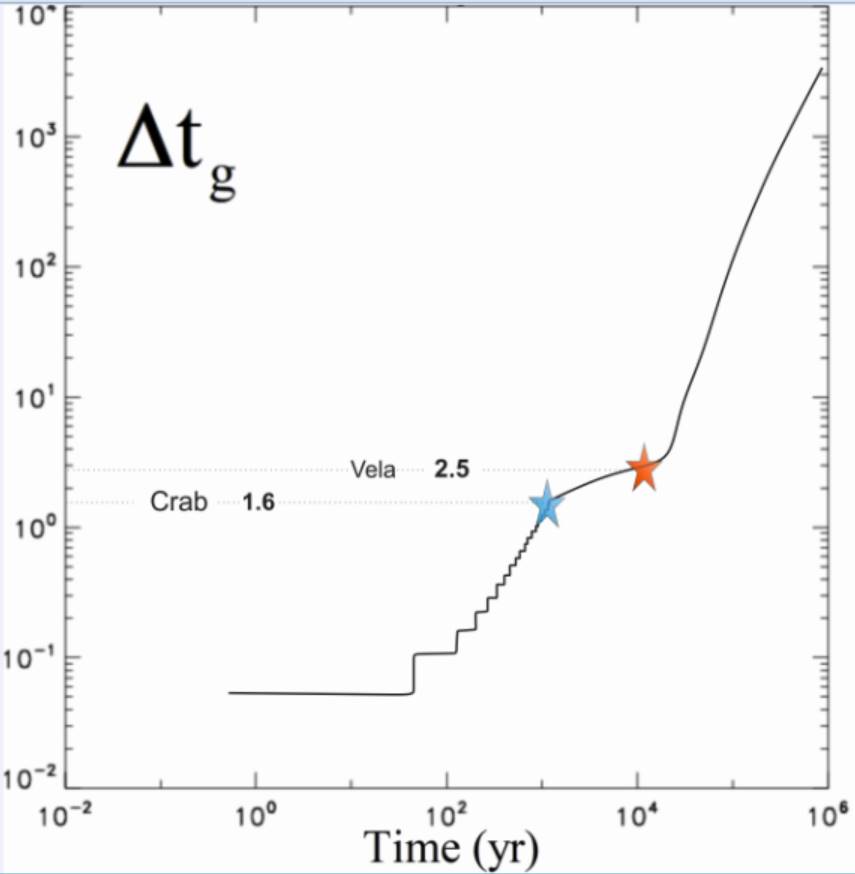}
}
\caption{\small  The elements of the sequence  $\Delta t_g$ versus cosmic time. Each element corresponds to the time
passage between two successive glitch events. The actual values that correspond to the Crab and Vela pulsars are shown
in blue and red stars. These time passages increase dramatically with time due to the decay of both magnetic field intensity
and rotational energy.
}  \label{dt_g}
\end{figure}
 \section{Solution procedure \& results}
\begin{figure}[htb]
\centering {\hspace*{-0.35cm}
\includegraphics*[angle=-0, width=7.75cm]{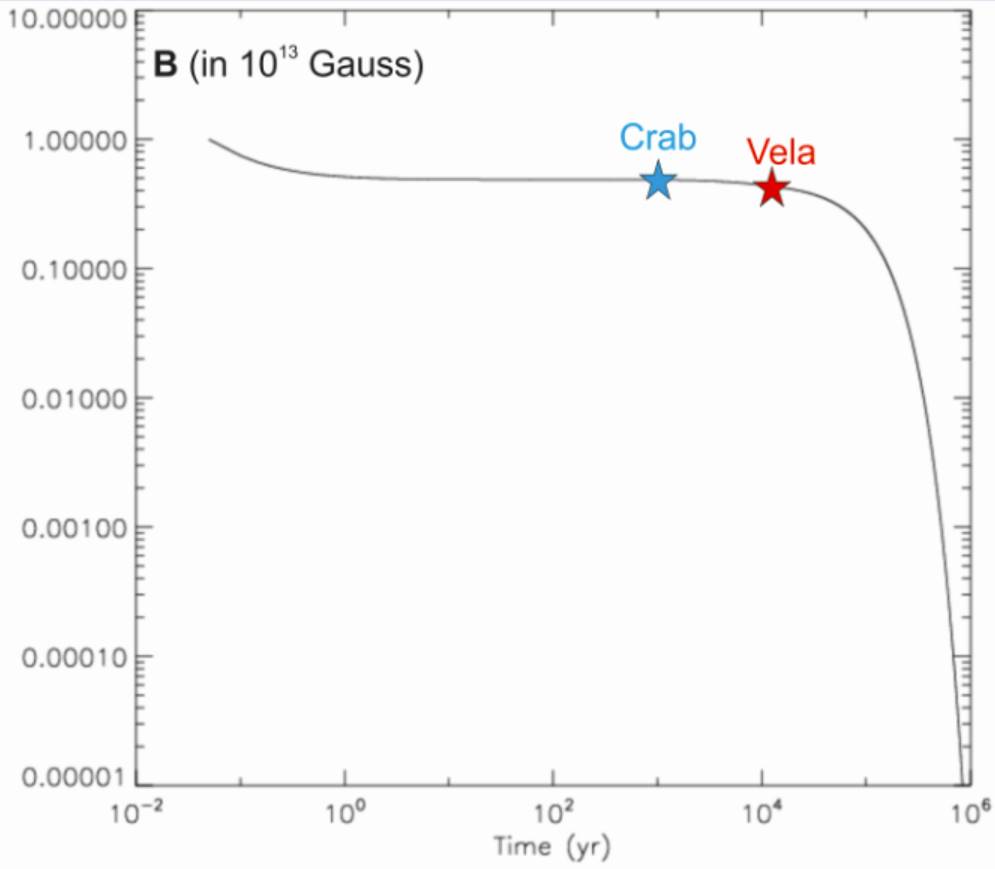}
}
\caption{\small The cosmic evolution of the magnetic field - B, of a newly born pulsar in units of $10^{13}$ G.  The superimposed
blue and red stars correspond to the current B of the Crab and Vela pulsars.
}  \label{BField}
\end{figure}

\begin{figure}[htb]
\centering {\hspace*{-0.35cm}
\includegraphics*[angle=-0, width=7.75cm]{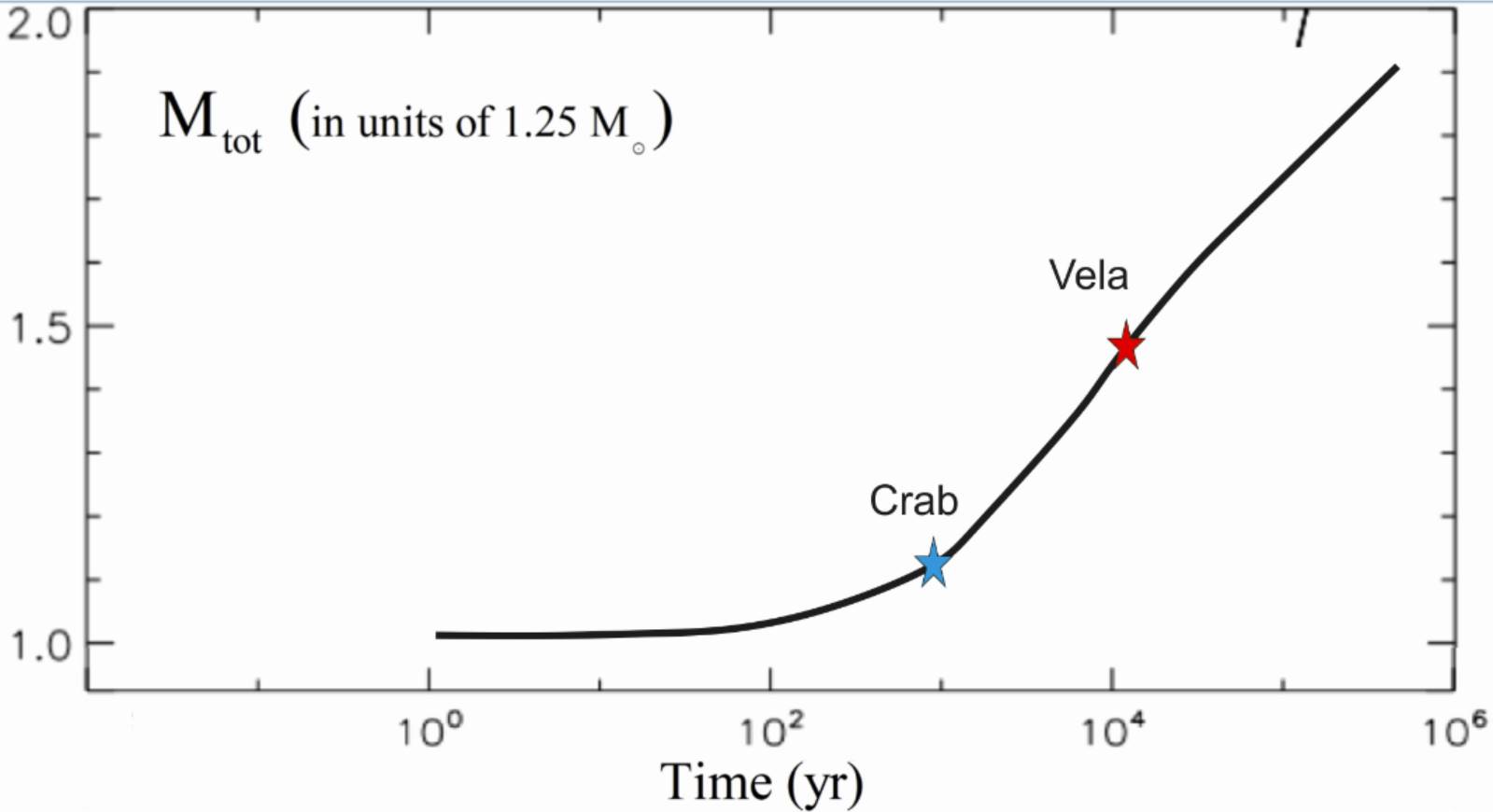}
}
\caption{\small The mass growth of a newly born pulsar having initially $M_0 = 1.25 \, \MSun$ and an embryonic SuSu-core of
$0.029 \, \MSun.$ After approximately 1000 yr the pulsar recover the mass of the Crab ($M_0 = 1.4 \, \MSun$)  and 10000 yr later
$M_0 = 1.8 \, \MSun$ that corresponds to the mass of the Vela. At the end of the luminous life time,
which lasts for  approximately 10 million years, the pulsar enters the  dark phase with a  total mass of $2.5~\MSun,$ which
corresponds to a maximally compact invisible dark energy object. The plotted mass here is in units of $M_0.$
}  \label{Mass_cosmic}
\end{figure}

 The set of equations consists of the TOV equation for modeling the compressible dissipative matter in the shell overlaying the
  incompressible gluon-quark superfluid core, whereas the latter is set to obey the zero-torque condition
  and to dynamically evolve according to  the Onsager-Feymann  equation (for further details see Sec. 2 and Eq. 10 in \cite{HujeiratGlitch18}).\\
 The global iteration loop is designed here to find the optimal values of the parameters: $ \alpha_0,\, \alpha_1,$   the elements
  of the sequence  ${\Delta\Omega^n_{cr}}$ and the decay rate of the magnetic field.  These values should fulfill the initial and final conditions,
  the currently observed values of the  time passages between two successive glitch-events $\Delta t_g$ both of the Crab and the Vela  pulsars,
  the current observed values of their magnetic fields masses.\\
\begin{figure}[H]
\centering {\hspace*{-0.35cm}
\includegraphics*[angle=-0, width=7.15cm]{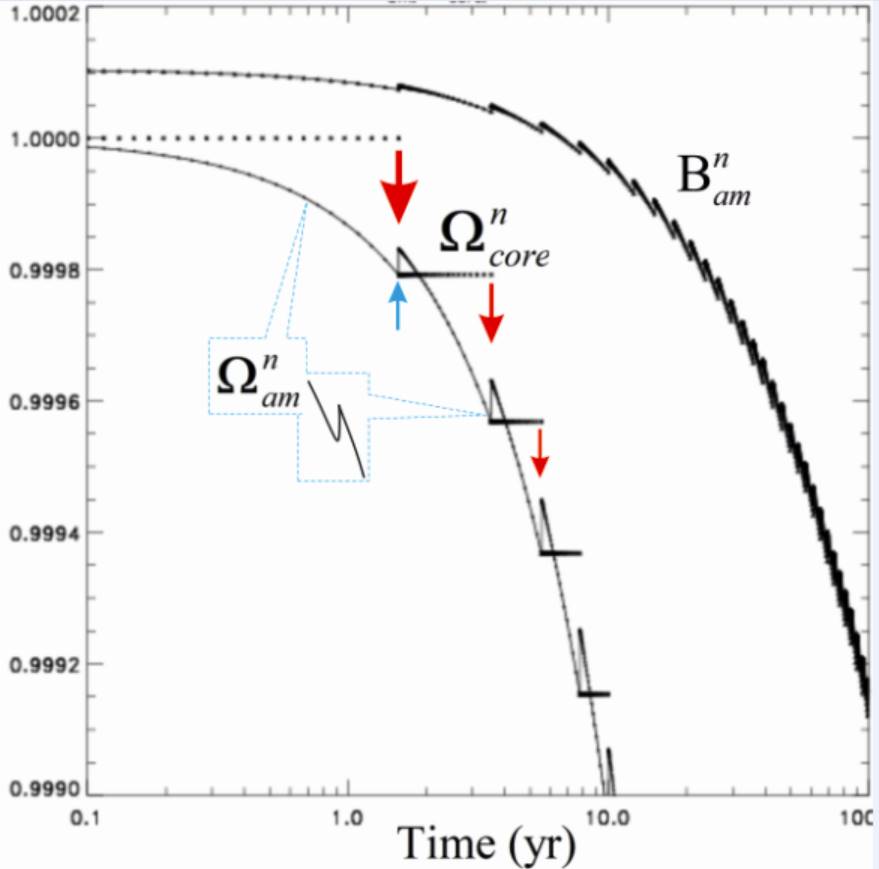}
}
\caption{\small
 The elements of the sequences $\Omega^n_c,\, \Omega^n_{am}$ and $B^n_{am}$ that corresponds to the  rotational frequencies
of the core, the ambient medium and of the magnetic field during the first 10 years. During a passage of time between two successive
glitches, the core rotates rigidly with a constant frequency $\Omega^n_c = \Omega^n_{core}$ (dotted line), whereas the ambient medium spin-down
in a continuous manner (solid line). During the glitch, the core spin-down abruptly, triggering a prompt spin-up of the ambient medium in the boundary layer between the rigid-body rotating core and the overlying differentially rotating medium.
The enhanced spin-up of the ambient medium in combination with the decreasing volume enclosing this matter gives rise to magnetic field $B_{am}$ which evolves in a similar discrete manner (dash-dot).
}  \label{Omg_jump}
\end{figure}
Indeed, our intensive computations reveal that optimal fitting may be achieved for $M_c(t=0) \approx 0.029 \,\MSun,$ a sequence of ${\Delta\Omega_{g}}^n,$ whose elements are shown in Fig.(\ref{domg_g}).  In Figs.(\ref{domg_cr},\ref{dt_g}) the optimal values of
 ${\Delta\Omega^n_{cr}}$ and ${\Delta\Omega^n_{g}}$ are shown versus cosmic time, whereas in Fig.(\ref{Omg_jump}) we show the time-development of the rotational frequencies of the core,  the ambient medium and of  the magnetic field during the first 10 to 100 years after the birth of the pulsar.
 The long-term evolution of the magnetic field and the growing mass of the core and of the entire object are shown in Figs.(\ref{BField} and \ref{Mass_cosmic}). Here the mass of the pulsar's core grows with time
to reach $0.15\, \MSun$  after 1000 years and reaches $0.55\, \MSun$  after 11000 years; hence reproducing the exact
total masses of $1.4\, \MSun$  for the Crab and $1.8\, \MSun$  for the Vela pulsars as revealed from observations. The relative ratio of inertia of both cores  reads:
$I^{Crab}/I^{Vela} \approx 3/20.$
\begin{figure}[htb]
\centering {\hspace*{-0.35cm}
\includegraphics*[angle=-0, width=7.15cm]{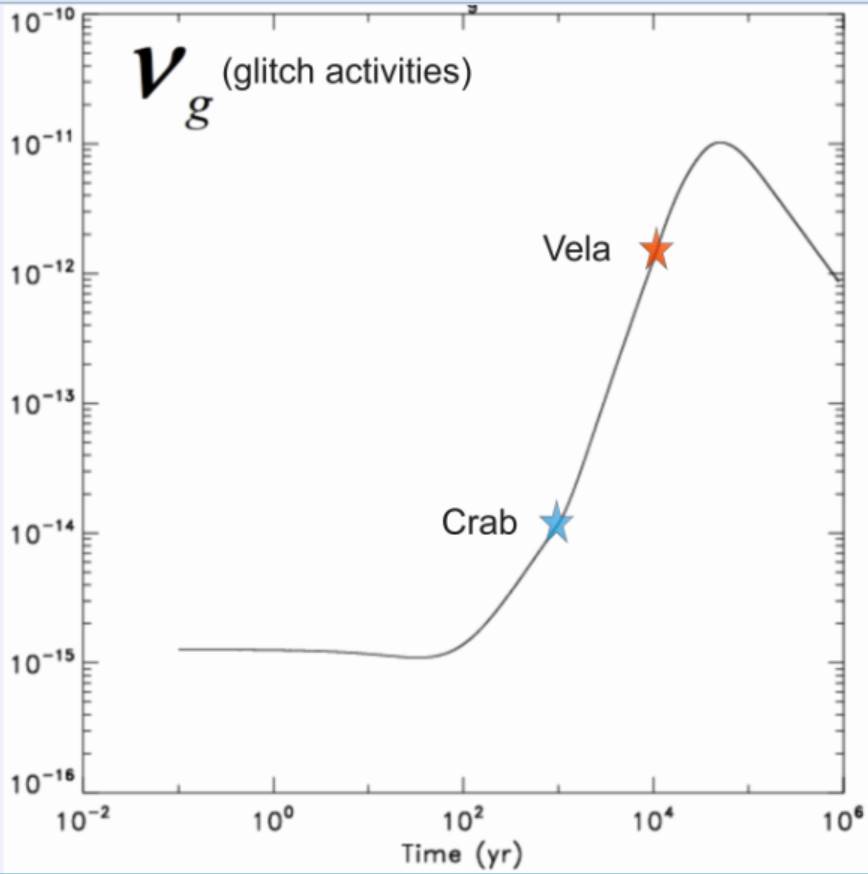}
}
\caption{\small The glitch activity of a newly born pulsars versus cosmic time. In the very early times, the pulsar underwent
millions of glitches, though the total ejected rotational energy was relatively very low. These activities start to be significant as the pulsar
ages and becomes maximally effective between 100 and 60000 years, followed by a  decreasing phase, during which
time-passages between successive glitches become increasingly longer.
}  \label{GActivity}
\end{figure}
\begin{figure}[htb]
\centering {\hspace*{-0.35cm}

\includegraphics*[angle=-0, width=5.5cm]{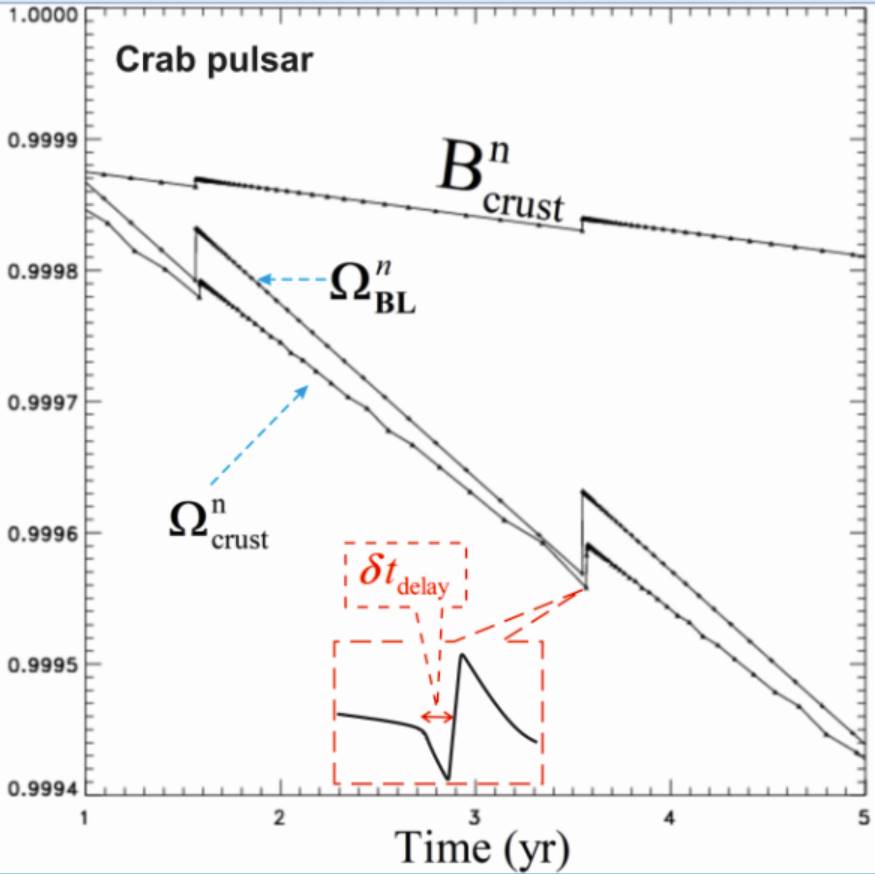}\\
\includegraphics*[angle=-0, width=3.5cm]{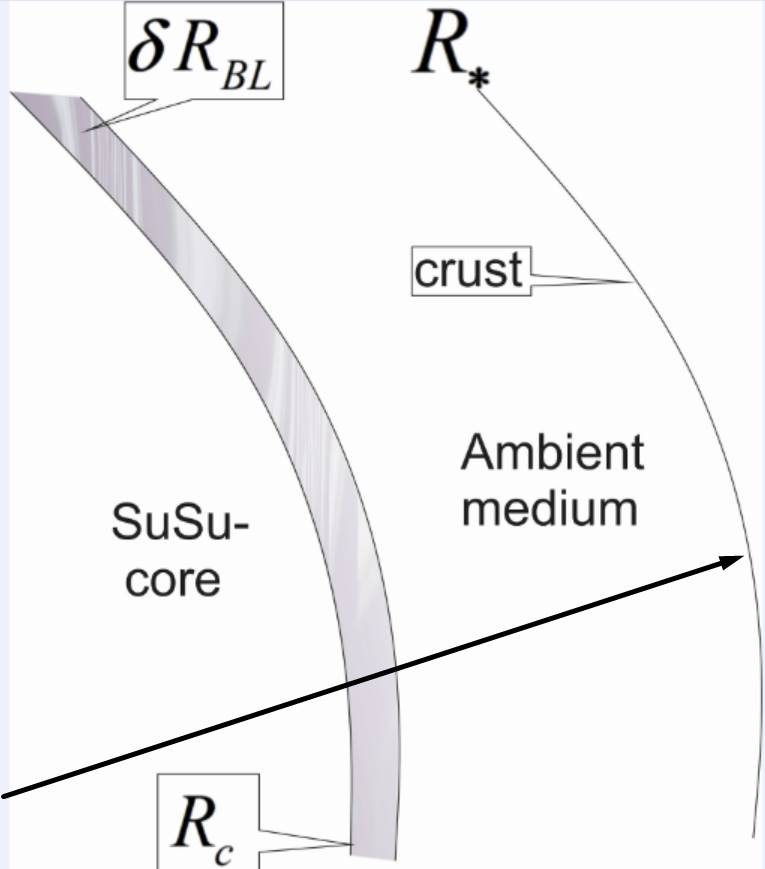}
}
\caption{\small Under and overshooting of $\Omega_{am}$ shortly before and immediately after a glitch event (top Figure).
At  a given instant of time, $t,$ the rotational frequency of the matter in the BL, $\Omega_{am},$ differs from that of the crust
$\Omega_{crust}.$ Due to their different locations (see lower panel),  the response of the crust to the dynamical changes of matter in the
 BL depends strongly on the speed of communication via magnetic fields (e.g. Alfven waves) and shear viscosity,
 which, under most astrophysical conditions, are considered to be different.
}  \label{UndershootingBL}
\end{figure}
Due to the incompressible, superfluid and supreconducting character of the core, the evolution of the magnetic field is solely
 connected to the dynamics of the ambient compressible and dissipative matter in  the shell as well as to its dimensions (see \cite{HujeiratLowMach2009} for further details on the physical aspects of compressibility of fluid flows).
 As the mass and dimension of the core grows with time, the surrounding shell must shrink. In this case,  conservation of the magnetic flux
  should strengthen the magnetic field intensity.  This interplay between the loss of magnetic energy due to loss of rotational energy
  and enhancement by conservation of magnetic flux in combination with dynamo action and other mechanisms, may clarify the
  the very weak decay of  magnetic field as pulsars evolve from the Crab to the Vela phase. Mathematically, let the magnetic energy
  in a shell of  a newly born pulsar be:
 \beq
 E_M = \int{\DD{B^2}{8 \pi}}dv \sim \DD{B^2}{6} (R^3_\star - R^3_c) ,
 \eeq
 where $R_\star $ denotes the pulsar's radius.
 Assuming $E_M$ to roughly decay as the rotational energy $ E_\Omega,$ then we obtain:
 \beq
 B^- = \alpha_\Omega   \Omega^2_{am} M^{1/2}_{am}\sqrt{\DD{ R^2_\star - R^2_c}{ R^3_\star - R^3_c}},
 \eeq
 where $ \alpha_\Omega$ is constant coefficient. \\
 On the other hand,  dynamo action in combination with magnetic flux conservation  and other enhancement mechanisms would
 contribute positively to the magnetic field,  that can, for simplicity absorbed in the term: $ B^+ = \alpha_B/ (R^2_\star - R^2_c).$  The
  coefficient $\alpha_B$ is set to ensure that the magnetic field remains in the very
 sub-equipartition regime. Hence  the interplay between magnetic loss and enhancement would yield an effective magnetic field
  that evolves according to:
 \beq
 B_{tot} = \DD{\alpha_B}{ (R^2_\star - R^2_c)}  -
 \alpha_\Omega   \Omega^2_{am} M^{1/2}_{am} \sqrt{\DD{ R^2_\star - R^2_c}{ R^3_\star - R^3_c}}.
 \eeq
 Consequently, our model predicts that the decreasing volume of the shell  enclosing the ambient medium in combination with dynamo action in the boundary layer could potentially be the  mechanism that keeps the decay of  magnetic fields  in pulsars extremely weak. \\

In fact our model predicts  the glitch activity of a newly born  pulsar, which  evolves
 into a Crab phase, followed by a Vela phase and finally by an  invisible phase, to be approximately two orders of magnitude larger than it was estimated by other models  (see Fig.  (\ref{GActivity} to be compared to \cite{Roy2012,Espinoza2011}). According to our model pulsar may undergo millions or up to billions of glitches during their luminous life time with passages of time between two successive glitch events that range
   from nanoseconds  in the very early time up to hundreds or even thousands of years toward the end of their luminous life times (see Fig. \ref{dt_g}).
   The vast difference  in the evolution of glitch activity between the two approaches here may be attributed to the strong non-uniformity of  time-duration between glitch events.

 Moreover, the model also predicts the occurrence of under- and overshootings that have been observed to accompany the glitch events in the Vela pulsar
 (see \cite{AshtonEtAl2019} and the references therein). In the case of the Vela, when the core expels certain number of vortices and moves to the next lower energy state, the enhanced rotational energy of the matter
 in the BL amplifies the magnetic field strength.  Due to the non-locality of magnetic fields\footnote{I.e. In the absence of magnetic monopoles}, this enhancement is communicated to the crust via Alfven waves, $V_A,$ whereas  the excess of  rotational energy is communicated via shear viscosity with an effective propagational velocity $V_{vis}.$ As these two speeds are generally  different  with $V_A > V_{vis}$ in most cases,  the time-delay in the arrival of communication enforces the crust to react differently. Specifically,   the arrival of magnetic enhancement prior to the rotational one leaves the crust subject to an enhanced magnetic braking and therefore to a stronger  reduction of its rotational frequency (see the top panel of Fig. \ref{UndershootingBL}). \\
 Indeed, in the case of Vela, the propagational speed  of Alfven waves may be esitmated to be of order $V_A \sim B/\surd{\rho} \approx 10^8$ cm/s. Hence the enhanced
 MFs in the BL would be communicate to the crust within $\delta \tau_{MF} = \Delta R/V_A= (R_\star - R_c) /V_A \approx 10^{-2}$ s.
 On the other hand, supplying the crust with rotational energy would proceed on the viscous time scale, which is estimated to be:
 $\delta \tau_{vis} = (\Delta R)^2/\nu_{vis}$ (see \cite{HujeiratTelemannAngMom2009} and the references therein). Under normal astrophysical conditions we
 may safely assume that $V_{vis}/V_A = \alpha_2 \ll 1$ and that the length scale, $ \ell_{vis},$
 over which viscous interaction occurs,  $ \ell_{vis}$  is much smaller than the width of the shell $\Delta R, $ or equevalently $ \ell_{vis} = \alpha_2 \Delta R, $ with $\alpha_2 \ll 1.$
  Therefore $\delta \tau_{Vis} = \Delta R/(\alpha_1 \alpha_2 V_A) \approx 10^{-2}/(\alpha_1 \alpha_2) \approx 1\,s, $
  where we reasonably set   $\alpha_1= \alpha_2=0.1.$
  Consequently, the observed undershooting most likely results due to  the time delay of the arrival of communications via magnetic fields and viscous
   torque, which amounts to $\delta \tau_{vis}/\delta \tau_{MF} = 1/ /(\alpha_1 \alpha_2) \approx 100.$
   On the other hand, the observed overshooting can be attributed to the case in which the viscous front transporting rotational energy from the BL
   outwards has reached the crust. As  $\delta \tau_{MF}\ll \delta \tau_{vis},$   at the end of $\delta \tau_{vis},$
   the magnetic field intensity in the BL  should have returned to values comparable or even lower than prior to the glitch event.

   Moreover, the observed order in which undershooting followed by overshooting is an indication for a
   time-delay in the arrival of communication resulting from $V_A > V_{vis} $ and  from the significant difference of the locations of the BL and the crust.   This order is expected to reverse if $V_A < V_{vis}.$

    In fact, the under- and overshooting here may indicate that MFs are insensitive to the momentary rotational frequency of the
    crust, but rather to the activity and dynamics of the matter in the BL. \\
   Extending this analysis to both the Crab and Vela pulsars,  the relative time-delays is expected to be: $\delta \tau^{Crab}_{vis}/\delta \tau^{Vela}_{vis} \sim (\Delta R^{Crab}/\Delta R^{Vela})^2 \approx 3.4$ or equivalently, the undershooting in the case of the Crab is expected
   to last 3.4 sec comapred to one second in the Vela case.

   Finally, although the physics is entirely different, the situation here is strikingly similar to action of the solar dynamo, which is considered to be located in the so-called
    tachcline between the rigid-body rotating core and the overlying convection zone \cite{Camenzind2007}.\\
{\bf{Acknowledgement}}
The calculations have been carried out using the computer cluster of the IWR, University of Heidelberg.
RS acknowledges the use of KAUS baseline research funds

 \end{document}